\documentclass[%
 reprint,
nofootinbib,
 amsmath,amssymb,
 aps,
]{revtex4-1}

\usepackage{graphicx}
\usepackage{dcolumn}
\usepackage{bm}
\usepackage{multirow}
\usepackage{hhline}
\usepackage{color}
\usepackage[normalem]{ulem}

\begin{document}

\preprint{APS/123-QED}

\title{Dynamics of Dengue with human and vector mobility }

\author{Murali Krishna Enduri}
\email{endurimuralikrishna@iitgn.ac.in}
 \author{Shivakumar Jolad}%
 \email{shiva.jolad@iitgn.ac.in}
\affiliation{
 Indian Institute of Technology Gandhinagar\\
 Chandkheda, Ahmedabad  380005, INDIA 
}

\date{\today}

\begin{abstract}
Dengue is a vector borne disease transmitted to humans by {\it{Aedes Aegypti}} mosquitoes carrying Dengue virus of different serotypes.  
Primarily an urban epidemic, Dengue exhibits complex spatial and temporal dynamics, influenced by many biological, human and environmental factors. 
In this work,  we study the Dengue spread for a single serotype including human mobility. We model the Dengue spreading by using PDE reaction-diffusion and 
stochastic Cellular Automata (CA) with human and vector dynamics and  analyze the spatial and temporal spreading of the disease using parameters from 
field studies.  Mosquito density data from Ahmedabad city serves as a proxy for climate data to our model. We predict the dynamics of Dengue incidence 
and compare it to the reported data on the prevalence of the disease from 2006-2012.  We find that for certain infection rates, CA model closely 
reproduces observed peaks and intensity. We have used statistical model of human mobility with exponential step length distribution to study mobility 
effects on Dengue spreading within the city. We find an interesting result that although human mobility makes the infection spread faster, there is an
apparent early suppression of the epidemic compared to immobile humans. The primary reason for decline is that mobility causes secondary and tertiary 
waves of infected individuals who recover in short time span and act as a barricade for the primary wave. 

\begin{description}
\item[Keywords]
Dengue, vector borne diseases, epidemics, cellular automata,  human mobility, reaction-diffusion.
\item[PACS numbers]
87.19.X-, 87.10.Mn , 87.23.Ge, 05.45.Tp, 07.05.Tp
\end{description}
\end{abstract}

\pacs{Valid PACS appear here}
\maketitle

\section{Introduction}

\begin{figure}[!t]
\centering
\includegraphics[width=3in]{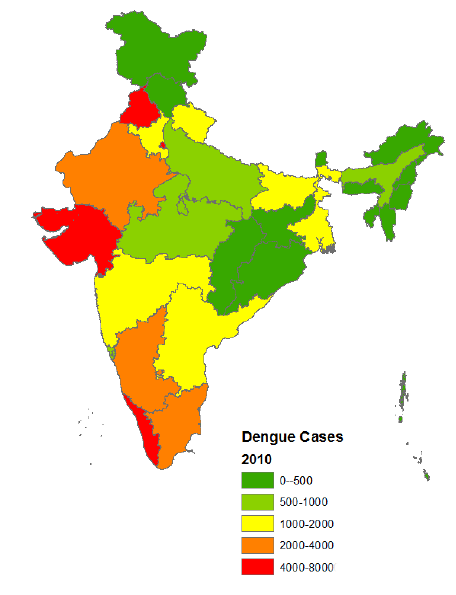}
\caption{Dengue cases in different states of India 2010. Data taken from National Vector Borne Disease Control Programme \cite{ncvrb2009:Online}.}
\label{fig:DIndiaCases}
\end{figure}

Dengue fever (DF) is a vector borne disease widely prevalent in tropical and subtropical regions in about 100 countries worldwide.
The World Health Organization (WHO) estimates that  over 2.5 billion people (40\% of the world population) are at the risk 
for Dengue and close to a million cases reported 2007 alone \cite{special2009dengue}.  Dengue is transmitted to humans mainly through {\em{Aedes Aegypti}} 
female mosquito bites carrying Dengue virus \cite{Nishiura2006}. Dengue fever (DF), Dengue hemorrhagic 
fever (DHF) and Dengue shock syndrome (DSS) are  different forms of Dengue infection, caused by four serotypes of Dengue virus (DENV:1-4)
\cite{special2009dengue}. The people who recover from one serotype can become permanently immune to it, but may not be immune to other serotypes.
Dengue is becoming a major public health concern in various South Asian and Latin American countries. In India, Dengue epidemic has spread to almost 
all the states and is posing a serious public health problem. In 2010 alone, 28000  cases were reported (see 
Fig. \ref{fig:DIndiaCases}). Dengue cases in India are massively under reported \cite{kakkar2012dengue}. A recent detailed study reveals that  
the actual cases are estimated to be more than 5.78 million between 2006-2012 \cite{Shepard2014, NYTDengue} in India.   Many Dengue infections may 
not produce severe symptoms, thereby evading early detection. At present, there is no effective vaccination or treatment for dengue. It is believed 
that any future dengue vaccination is imperfect \cite{bhamarapravati2000live},  and may not offer protection against all serotypes.  The only known 
effective way to prevent dengue outbreak is to devise  vector control strategies and minimize vector-human transmission. A sound understanding of the 
spatial and temporal dynamics of the Dengue can help in devising strategies for containing the spread urban populations.  

Numerous human, biological, social and environmental factors affect the transmission of Dengue \cite{MacieldeFreitas2011452,kuno1995review,Adams2010}.  
Several mathematical  models have been  proposed (see \cite{Nishiura2006,derouich2006dengue,Andraud2012}  for reviews) for studying the Dengue. Many of 
these are the compartmental ordinary differential equation (ODE) models \cite{anderson1991infectious, bailey1975mathematical}, which  divide the human 
population into Susceptible, Exposed, Infected,  and Recovered (SEIR) groups; and vectors into Susceptible, Exposed, Infected  (SEI) 
groups \cite{newton1992model}, and studying their temporal dynamics. These ODE models are essentially mean field models which neglect the spatial 
patterns of the spread of Dengue.  Attempts at spatial modeling of Dengue 
includes those based on spatial data mapping and statistical analysis \cite{Bhandari2008, Bohra2001,peterson2005time,rotela2007space}, reaction-diffusion  
partial differential equation (PDE) with vector or larval mobility \cite{Tran2006,Maidana2008}, Individual based models (IBM) on a 
grid \cite{bian2004conceptual, Otero2008, otero2011modeling, Barmak2011} 
and Cellular Automata  with vector mobility and simplified models of human mobility \cite{de2011modeling,santos2009periodic}.  In this work, 
we first report results of SEIR-SIR reaction-diffusion PDE  model as a reference to our full scale study of spatio-temporal dynamics through Cellular 
Automata.

Human mobility, especially of the infected individuals can create multiple Dengue waves resulting in substantial deviation  from mean field results.  
However, the current approaches to study Dengue spreading with human mobility have been restricted to simple methods such as movement with fixed step size, 
introducing a global field altering transition probabilities \cite{de2011modeling,santos2009periodic} and metapopulation with static vectors \cite{Adams2009}.
It has been studied at multiple scales such as house to house mobility \cite{Stoddard2013}, rural-urban daily commuters \cite{mpolya2014epidemic}  
and Dengue spread across countries \cite{wichmann2004dengue}.   In this work we use {\it stochastic} Cellular Automata (CA) approach 
(closely following \cite{de2011modeling}), with realistic models  of human mobility patterns derived from statistical studies of human 
movements observed through the circulation of currency notes, tracking of phone calls through Cellular towers, and location based social networks 
such as Foursquare.  These works have shown varied patterns such as L\'evy flight (Brockman {\it et al.} \cite{Brockmann2006,Brockmann2007}), 
truncated L\'evy flight (Gonsalez {\it et al.}  \cite{Gonzalez2008}),  exponential distribution in intra urban movements ( Liang {\it et al.} 
\cite{liang2013unraveling} and Noulas {\it et al.} \cite{noulas2012tale}). The differences arise possibly due to the difference in scale and 
resolution of study (large distance, intra city movements, mobile tower coverage etc.) and the methodology used. Since Dengue is primarily an 
urban disease, we focus on the spread within an urban area and use exponential tail \cite{liang2013unraveling} distribution for studying human 
mobility affects on spread of Dengue.  

\begin{figure}[ht!]
\centering
\includegraphics[width=3in]{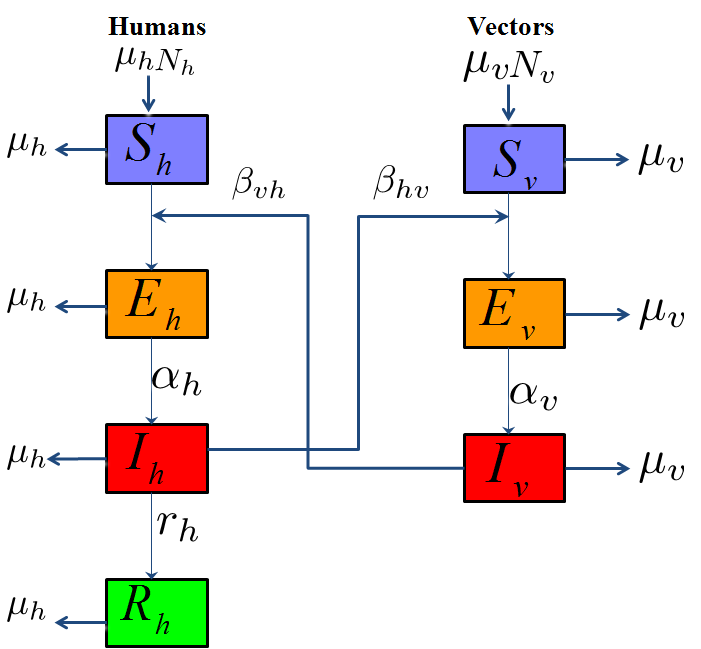}
\caption{SEIR-SEI model of human-vector interactions.}
\label{fig:SEIRSEI}
\end{figure}
 
\section{Model Formulation}

In this study, we use the standard compartmental model to divide the humans into SEIR and vector population into SEI groups 
(see \cite{Nishiura2006, Andraud2012} for review). Only infected vectors and infected humans can transmit the dengue virus to 
susceptible population (See Fig. \ref{fig:SEIRSEI} for illustration). Exposed population is infected with the vector but {\it not infectious} 
(i.e. they cannot transmit the Dengue virus).  In Fig. \ref{fig:SEIRSEI}, we show the flow diagram of SEIR-SEI model. The temporal evolution of 
the corresponding ordinary differential equations, their stationary states and stability conditions have been investigated and reported in many 
works (see for Eg: \cite{Nishiura2006,Pongsumpun2008}).

The compartmental ODE models assumes homogeneous mixing of population with no spatial information is encoded in it. Spatial approaches are needed 
to analyze local and global dynamics and present a more realistic picture of the disease propagation. Here, we focus on the spatial approaches to 
modeling Dengue based on reaction diffusion equations.

\begin{table*}[ht!]
\centering
\begin{normalsize}
    \begin{tabular}{|l|l|l|}
    \hline
    {\bf Parameter}    & {\bf Description}                                                  & {\bf Value}              \\ \hline
   $\mu_h$ & Death rate of humans                                         & 0.0000391 per day  \\ \hline
    $\mu_v$ & Death rate of vectors                                        & 0.07142 per day    \\ \hline
    $\beta_{hv}$ & Infection rate  from human to vector         & 0.00008            \\ \hline
    $\beta_{vh}$ & Infection rate from vector to human         & 0.00005            \\ \hline
    $r_h$ & Recovery rate of human population                            & 0.07142 per day    \\ \hline
    $\alpha_h$ & Rate at which exposed human change to be infected  human    & 0.2 per day        \\ \hline
    $\alpha_v$ & Rate at which exposed vector change to be infected vector  & 0.1 per day        \\ \hline
    $F$        & Flight range                                                 & 7                  \\ \hline
    $D_v$        & Diffusion Coefficient                                        & 0.2                \\ \hline
    \end{tabular}
\caption{Table of parameters reaction-diffusion model (see Eq.\ref{eq1} and \ref{eq2}) \cite{Pongsumpun2008}. }
\end{normalsize}
\label{tab:ParamPDE}
\end{table*}

\begin{figure*}[ht!]
\centering
 \includegraphics[trim=1.8cm 21.0cm 1cm 2.5cm, clip=true, scale=0.9]{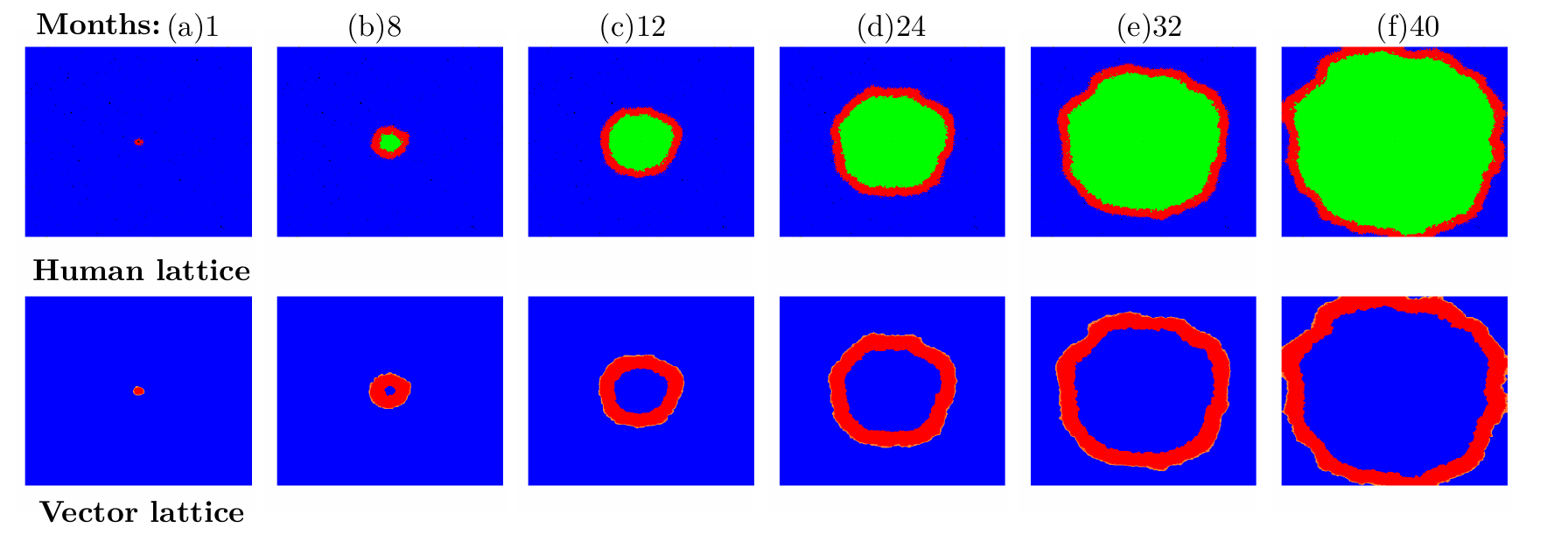}
 \caption{(color online) Spatial temporal spread of dengue with reaction-diffusion equations with only vector mobility.
 Grid size is $500\times 500$, $D_h=0$. The upper row corresponds to human grid and the bottom row to vector grid.
 coloring scheme depend on the density of S(blue)-E(orange)-I(red)-R(green) and it is described in Sec. \ref{sec2A}. }
 \label{fig:PDEVM}
\end{figure*}

\subsection{Reaction-diffusion approach}\label{sec2A}

 In vector borne diseases, spatial spreading  is possible only when there is mobility of vectors, humans or both. 
Vectors, especially \textit{A. Aegepti} rarely fly long distances by itself, and hence their mobility can be modeled a 
diffusion process  \cite{Tran2006,Maidana2008}.  Their long distance mobility requires external drivers such as wind, vehicles, 
ships \cite{Otero2008} and will not be considered here.  Human mobility is more complex with many works showing different statistical 
pattern of step length depending upon the scale, resolution and range of the study . A detailed discussions will be provided in section \ref{subsec:HLM}.  

Reaction-diffusion (RD) equations  are formed by adding diffusion terms to the reaction equations of the compartmental ODEs. 
Here, we model human mobility as a diffusive process and try to understand how the relative strengths of diffusion coefficients 
$D_h/D_v$ affect the spatio-temporal spread of Dengue. The RD equations for the humans in stages  $\{S,E,I,R\}$ is given by:  
\begin{eqnarray}\label{eq1}
 \frac{\partial s_h}{\partial t} &=&D_h \nabla^2 s_h +\mu_h n_h-F^2\beta_{vh} n_h i_v-\mu_h s_h \nonumber \\
 \frac{\partial e_h}{\partial t} &=&D_h \nabla^2 e_h+F^2\beta_{vh}e_h i_v-\alpha_h e_h-\mu_h e_h \nonumber \\
 \frac{\partial i_h}{\partial t} &=&D_h \nabla^2 i_h+\alpha_h e_h-r i_h-\mu_h i_h \nonumber \\
 \frac{\partial r_h}{\partial t} &=&D_h \nabla^2 r_h+r i_h-\mu_h r_h .
\end{eqnarray}
For the vectors, the modified equations with the diffusion term  and flight  range $F$ are : 
\begin{eqnarray}\label{eq2}
 \frac{\partial s_v}{\partial t}&=& D_v \nabla^2 s_v  +\mu_v n_v  -F^2\beta_{hv} s_h i_v-\mu_v s_v \nonumber \\
 \frac{\partial e_v}{\partial t}\!&=& D_v \nabla^2 e_v  + F^2\beta_{hv} s_v i_h   -\alpha_v e_v-\mu_v e_v \nonumber \\
 \frac{\partial i_v}{\partial t}\!&=& D_v \nabla^2 i_v + \alpha_v e_v-r i_h-\mu_vi_v , 
\end{eqnarray}
with $n_h=s_h+e_h+i_h+r_h, n_v=s_v+e_v+i_v$. Here $\mu_h,\mu_v$ represent human and vector mortality (death) rates; 
$\beta_{vh},\beta_{hv}$ represent the vector to human and human to vector infection  rates; $\alpha_h, \alpha_v$ are the 
transition rates from exposed to infected states in humans and vectors; $r$ is the recovery rate.   The diffusion term 
$\nabla^2= \frac{\partial^2}{\partial x^2} +\frac{\partial^2 }{\partial y^2}$ is the 2D Laplacian operator. 
The flight range $F$ indicates the maximum number of cells, vectors can hop in at a time. The above equations assume that 
the total population of humans and vectors is conserved. The death rate for humans (about 1/(75yrs)) is much lower than that of 
vectors (about 1/14 days), hence the dynamical time scales vary widely. All the parameter values are shown in Table \ref{tab:ParamPDE}.

\begin{figure*}[ht!]
\centering
 \includegraphics[scale=0.9]{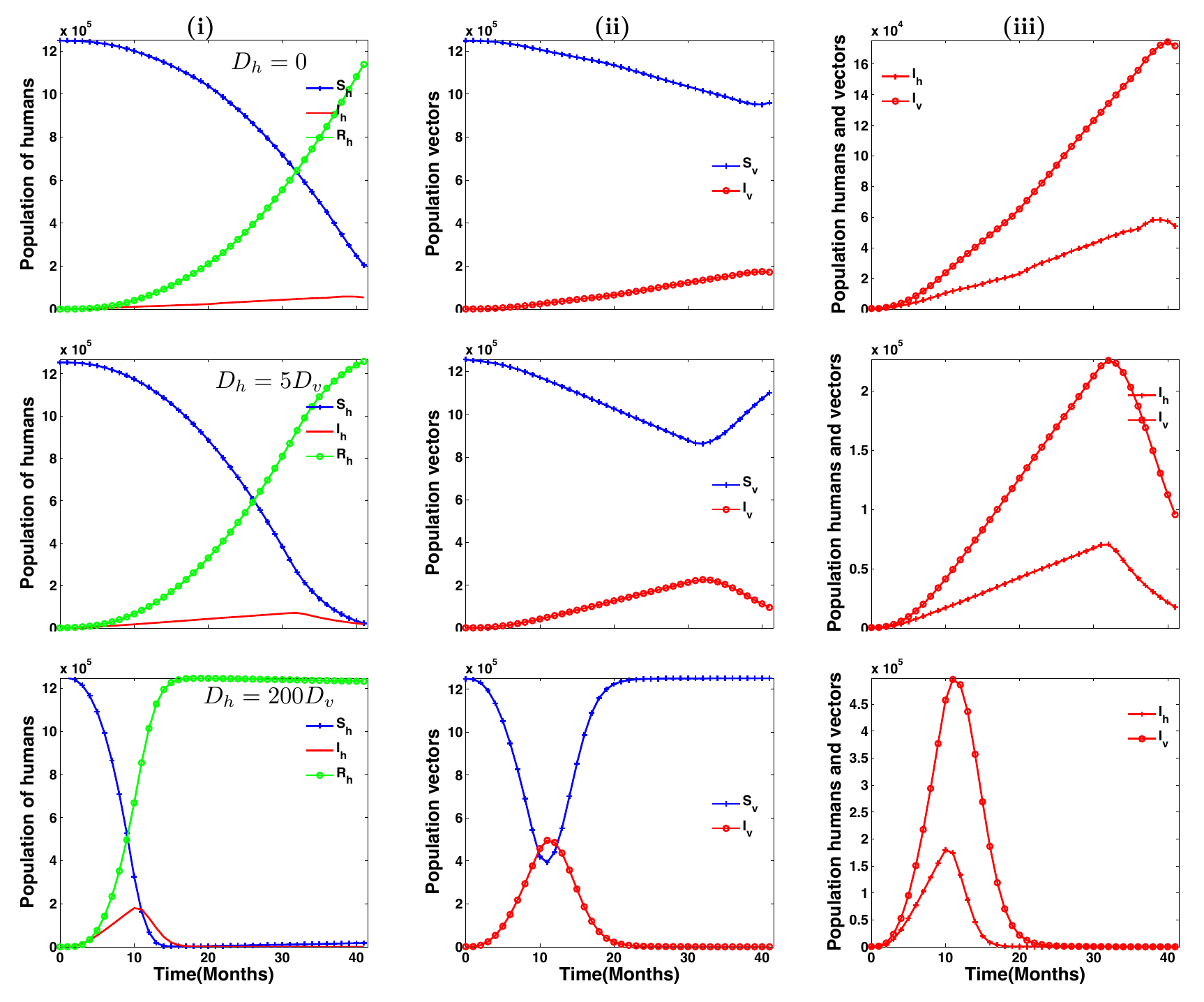}
  \caption{(color online) Temporal dynamics of Dengue reaction-diffusion model described in Eq. \ref{eq1} and \ref{eq2}. Columns:  (i) SIR for humans, (ii) S-I for vectors (iii) Infected humans and vectors. Rows: diffusion rates 
  1) $D_h=0$ 2) $D_h= 5D_v$ 3) $D_h= 200 D_v$, with $D_v=0.2$. }
  \label{fig:pde_h_v}
\end{figure*}

The spatial spread of Dengue in humans and vectors on a $500\times 500$ mesh grid with immobile humans ($D_h=0$) 
are shown at various intervals in Fig. \ref{fig:PDEVM}.  The top (bottom) row shows the density of humans (vectors) in 
various stages of the disease. Initially (at t=0), all humans are susceptible and all vectors, except the ones at the center are susceptible. 
Only the central cell vectors are initially infected with Dengue.   The coloring scheme for cells in the spatial grid are chosen according to 
the following rules: If the susceptible (recovered) density exceeds  0.5, the cell color is blue (green), or if the infected (exposed) density 
exceeds 0.25, it is painted red (orange). The same scheme is used in Cellular Automata later.

The overall pattern of the Dengue spread appears like a circular wave spreading outwards. This is due to the assumption of homogeneous and 
isotropic distribution of population, and diffusion constants being same in all directions. We observe that infection in the vector grid 
spreads faster than the human grid. We should note that the death rate of vectors is much higher than that of humans, but the incubation period is 
longer in vectors than humans ($\alpha_h>\alpha_v$). Hence we can observe the exposed vector population as a faint orange ring leading the infection. 

The temporal variations of humans and vectors in different disease states for $D_h=\{0,$ $5D_v$, $200D_v\}$ are shown in Fig.\ref{fig:pde_h_v}. 
The top row corresponds to immobile humans ($D_h=0$). It exhibits typical SIR behavior up to 40 months. Afterwards the wave hits boundary 
(see Fig. \ref{fig:PDEVM}-f), halting the infection spread and with time, and everyone recovers.  In the middle  and bottom panel, we set human 
mobility to be same or higher order than the vectors by choosing $D_h=5D_v$ and $D_h=200D_v$ respectively. Comparing these rows, we clearly see 
that the inclusion of human mobility causes early raise in the infected and then decline (once the wave hits the boundary and every one recovers 
after the incubation period). This makes it appear that in the long run that inclusion of human mobility shows lesser level of infection level 
than that for the immobile case.

\subsection{Stochastic Cellular Automata}
\label{subsec:CA}

\begin{figure*}[t!]
\centering
 \includegraphics[trim=1.8cm 21cm 1cm 2cm, clip=true, scale=0.8]{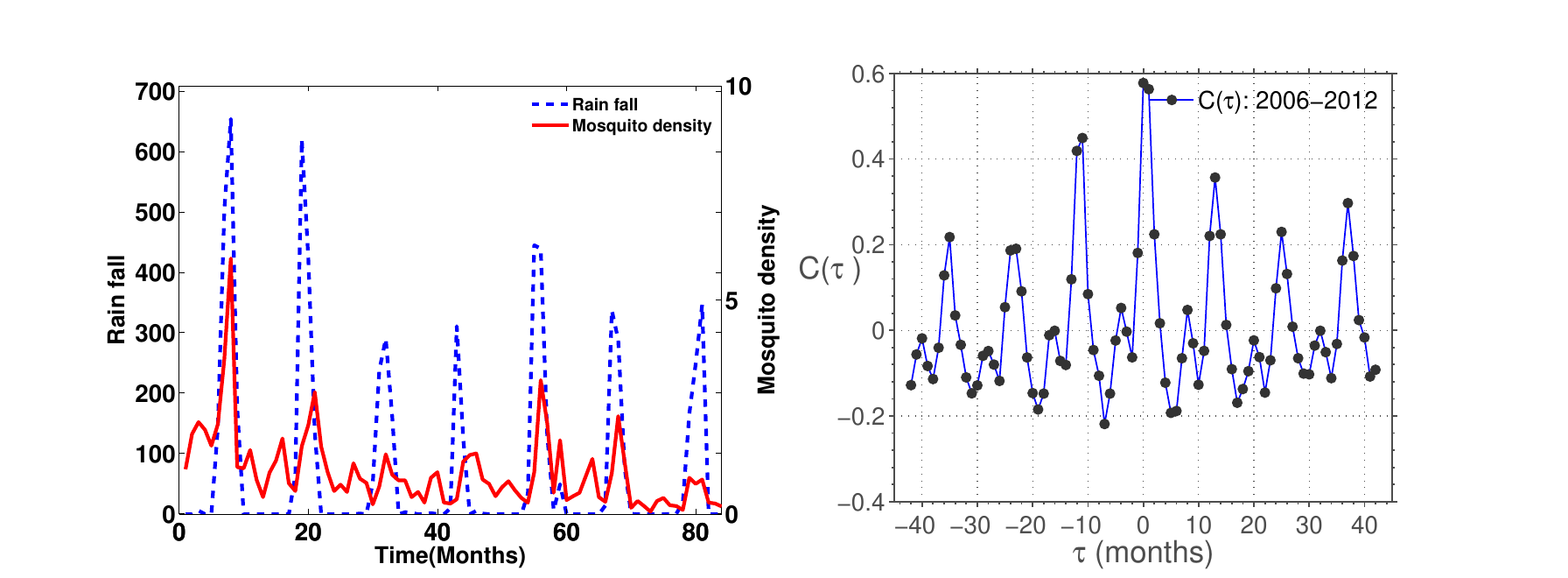}
\caption{(a) {\it A. Aegepti } mosquito population and rainfall  in Ahmedabad from 2006-2012 (source:  Ahmedabad Municipal Corporation) 
(b) Corresponding cross correlation of rainfall and mosquito density.}
\label{fig:MosqDenisty}
\end{figure*}

In Cellular Automata, space is divided into discrete grid and a finite state machine is assumed to operate at each grid point. 
Time is discrete, and at every step each cell is in one of a finite set of possible states. The transition probabilities depend only 
on the present state of the cell and its neighbors \cite{sarkar2000brief}.  

We model the spatial dynamics of Dengue using the stochastic Cellular Automata formalism by placing  human and vector ``agents" on  
two overlapping lattices and different disease states of  these agents \cite{de2011modeling}. We impose transition probabilities between 
the disease states based on field studies. For vectors we impose diffusive pattern, and for humans we use the exponential distribution of 
step length as described before.  Climatic factors such as rainfall and temperature strongly affect the vector population. We have used mosquito 
density data of Ahmedabad city in mid-western India (with population of about 6 million) from 2006-2012 as proxy for climatic factors. We compare 
the results of simulation of Dengue incidences with the recorded cases in Ahmedabad from 2006-2012. Our primary focus though will be on understanding 
how mobility factors affect the spatio-temporal pattern of Dengue. 

We choose the human ($H$) and vector ($V$) lattice to be of the same size. Each cell $H_{ij}$ of human layer can have multiple agents (humans) in 
different disease states \{S,E,I,R\}, which can change with time. For example consider a cell $H_{23}$ with 5 humans with $\{3S,2E,0I,0R\}$. In the 
next time step, it can be $\{2S,2E,1I,0R\}$. Similarly, in vector cell, mosquitoes can have states \{S,E,I\}, which can change with time. Throughout 
our simulations we take each time step to be one day.  In Table .\ref{tab:ParamTable}, we summarize the main  parameters used in our CA simulations.

\subsubsection{Vector population and mobility }
\label{subsec:vectors}
In each cell of the vector lattice $V_{ij}$, mosquitoes are distributed according to the Poisson distribution ${\cal P}(\lambda_v)$, where  
$\lambda_v$ is the average density of the {\it A.Aegepti} mosquitoes in  the cell. We use the periodic (monthly) data, denoted by  
\{$N_{m_1},N_{m_2},\dots N_{m_{n}}$\}, on the {\it Aedes} mosquito density available for Ahmedabad city in India from 2006-2012 
(plotted in Fig. \ref{fig:MosqDenisty} (a)) for simulations. The vector population is strongly correlated with the climatic factor, 
especially rainfall. In Fig. \ref{fig:MosqDenisty} (b) we show correlation of \textit{Aedes} density with the rainfall during the same 
period. We observe peaks at periodic intervals, with the dominant one at multiples of 12 months.

At the start of the simulation, in each cell the mean occupation is assumed to be $\lambda_v=N_m$, the measured average per room density of mosquitoes.   To interpolate between the monthly data, we assume an exponential growth $n(t)=n(t_0)e^{r(t-t_0)}$ between month $m_k$ and $m_{k+1}$, with growth rate $r=\frac{1}{30}\ln(N_{m_{k+1}}/N_{m_k})$, $n(t_0)=N_{m_k}$ and $t_0\le t < t_0+30$. We denote the  population in each cell  from simulations be $n_s(t)$ and assign new births/deaths depending whether   $\delta n=n(t)-n_s(t)$ is positive or negative. Although vertical transmission of Dengue virus from mother to newborn mosquitoes  is recorded, studies have shown that it does not affect long term virus persistence, due to low vertical infection inefficiencies \cite{Adams2010}. Hence we set the new born mosquitoes to be \emph{susceptible}. If $\delta n >0$, we generate new susceptible mosquitoes with Poisson distribution ${\cal P}(\delta n)$. If $\delta n<0$, we calculate the a random number $m={\cal P}(-\delta n)$ and kill them in that cell.

{\bf Vector Mortality:} \textit{A. Aegepti}'s life span is short with average of 22 days and  maximum of 45 days \cite{MacieldeFreitas2011452}.  Its death rate is not significantly affected by whether it is carrying the Dengue virus or not.  Here we set the vector mortality rate to be constant  $\mu_v$ per day.

{\bf Vector Mobility}: We model the vector mobility by hopping of mosquitoes between cells. At each time step, 50\% of the mosquitoes 
decide to move out of the cell. The probability of hopping to a Moore neighborhood\footnote{It consists of cells in a square neighborhood at an edge cell distance of $r$ from the center. There are $(2r+1)^2$ cells in such a ring \cite{sarkar2000brief}.}  of range $r$  is $1/2^r$, with $r=1,2,\dots,\infty$. Within this range, the probability of choosing any cell is  uniform: $p=1/((2r+1)^2-1)$.  If we take $a$ to be the width of each cell, the average distance along sides is $a \sum_{r=1}^{\infty} r/2^r= 2a$ (flight range of $f_r=25m$), and along the diagonal is $2a\sqrt{2}$.

\subsubsection{Humans Lattice and Mobility}
\label{subsec:HLM}
In each cell, human occupation is chosen to be Poisson distribution ${\cal P}(\lambda)$ with mean $\lambda=5$. 
The birth rate is balanced with the death rate to keep the population constant. Human mortality rate ($1/70yrs$) 
is much smaller to compared to the vector and hence does not account for significant deaths during the simulation time. We also assume 
that the death rate due to Dengue is zero.

Human mobility is primarily responsible for carrying communicable diseases to large distances, both within the city and across the cities and countries \cite{wichmann2004dengue}. Vector borne diseases can be spread by both human and vector mobility. A recent study with smaller 
population on $20\times 20$ blocks, with a network based link length  distribution following L\'evy flight pattern shows that human mobility 
strongly enhances the infection dispersal in vector borne  diseases \cite{Barmak2011}.   Typically human mobility cause the virus to carry disease 
across different regions in much shorter time period than the vectors. Mobility of infected humans or vectors can create multiple waves of diseases 
at different locations. In SIR/SEIR model, mobility can also lead to depletion of infected in a particular region and hence local reduction in the 
transmission rates. Such competing forces call for a careful study of the mobility effects on the spread of vector borne diseases.

 Many factors influence mobility within a city  such as population density, transportation networks, traffic patterns and varying economic activity  in different localities (home, work, school, shops, hospitals etc.). Modeling spatial human mobility in cities (especially in India) in the absence of credible data is a daunting task. Several ingenious methods have been used in the past to study the statistical patterns of human mobility.  Tracking of currency notes yielded a scale free L\'evy flight pattern \cite{Brockmann2006} $P(\Delta r)\sim (\Delta r)^{-(1+\beta)}$ across large scale. Later, a study based on the trajectory of 100,000 anonymised mobile phone users \cite{Gonzalez2008} in US showed that the step length distribution  behaves like a truncated L\'evy flight $P(\delta r)=\frac{A}{(\delta r+ \delta r_0)^\beta}\exp(-\delta  r/\kappa)$. 
 
The advantage of these methods is that statistical patterns are robust and does not critically depend on the variations in transportation networks or population density. Dengue is primarily an urban disease, where mobility within the city is more important than across. Study of intra-urban mobility received special attention in the recent years. Liang {\it et al.} \cite{liang2013unraveling} 
have produced a strong evidence of exponential distribution in intra-urban movements.  It is supported by a recent study by  Noulas {\it et al.}, where they explore intra city movements using location data by social networking site Foursquare \cite{noulas2012tale}. In this work, we choose exponential step length model $ P(\Delta r)=\lambda e^{-\lambda\Delta r}$ for studying human mobility.  Liang {\it et al.} found the mobility exponent $\lambda$ is not universal, and varies like 0.08 $km^{-1}$ in Los Angeles to 0.22 $km^{-1}$ in Beijing \cite{liang2013unraveling}. 
In our model world, we choose the distribution in terms of characteristic step length $w$ to be
 \begin{equation}
P(l)=\frac{1}{w}e^{-l/w},
\label{Eq:expstepcell} 
 \end{equation}
where $l$ is step length scaled to the lattice size.

In this work, we restrict the mobility to S, E and R population. Infected  are immobile as they are at rest/hospitalized. With a 50\% probability of S, E or R type people move from their current cell following above distribution. Once a person decides to move, a random number following exponential distribution (as in Eq. \ref{Eq:expstepcell}) is drawn and step length is determined. The angle is chosen from a uniform distribution $U(0,2\pi)$. The cell which contains the location $(\delta r, \theta)$  from the current point is chosen as the 
destination cell. If the range is outside the CA boundary, periodic boundary conditions bring the person back into the CA world. This ensures that net migration in and out of CA is zero.  

\begin{table}[t!]
\centering
  \begin{tabular}{|l|m{3.5cm}|m{3.2cm}|}
 \hline
    Parameter & Description & Values \\ \hline
    $\tau_{hE}$  & The duration for exposed human to become Infected & Uniform distribution: 4-7 days \cite{halstead2008dengue} \\ \hline
    $\tau_{vE}$  & Infected vectors  incubation period & Uniform distribution: 8 to 12 days  \cite{special2009dengue, gubler2001climate} \\ \hline
    $f_r$        & Maximum flight range of vectors  & 25 m \cite{kuno1995review} \\ \hline
    $B_r$        & Biting rate of vectors  & 0 to 2/day \cite{de2011modeling} \\ \hline
    $\beta_{h \nu }$,  $\beta_{\nu h}$        & Human to vector and vector to human transmission rates  & \{0.15, 0.4, 0.6, 0.9\}  \\ \hline
    \end{tabular}
   \caption{Value of parameters for stochastic Cellular Automata simulations}
   \label{tab:ParamTable}
 \end{table}

\begin{figure}[b!]
\centering
\includegraphics[width=3in]{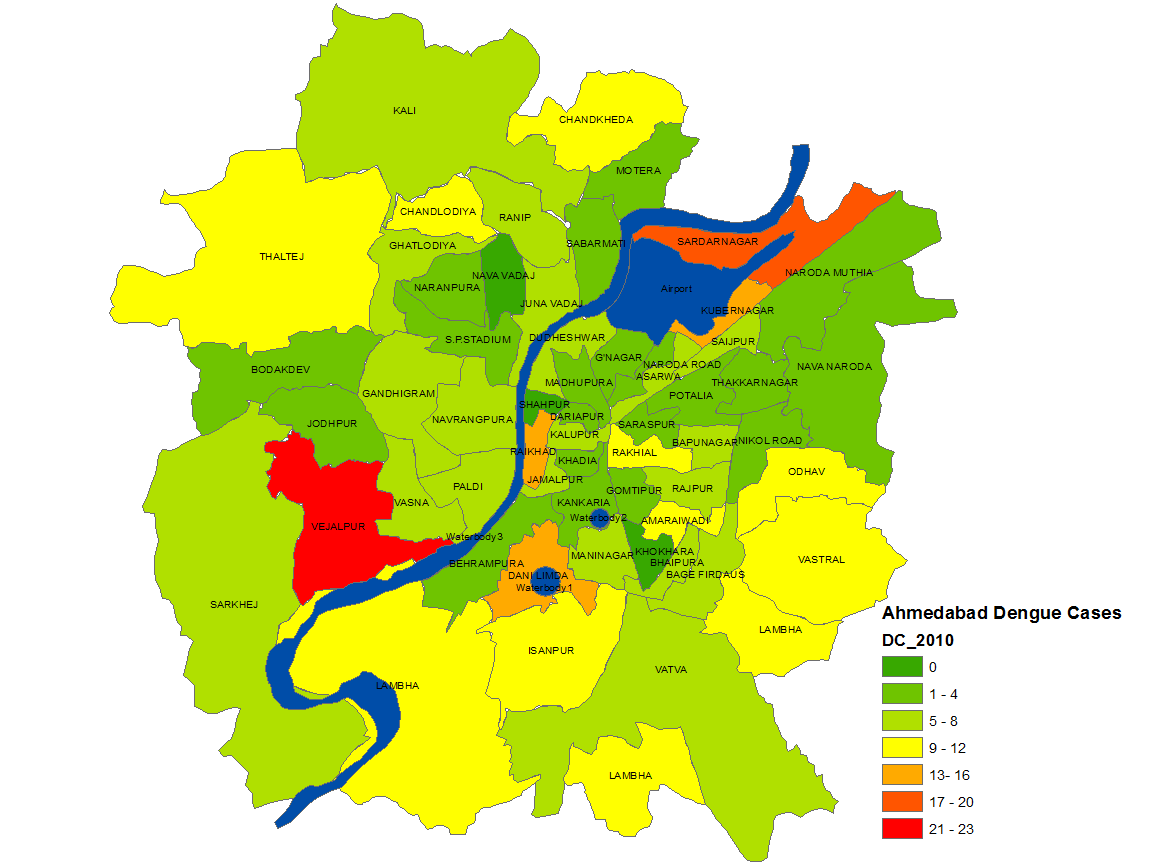}
\caption{Dengue cases in Ahmedabad 2010 (source: Ahmedabad Municipal Corporation). }
\label{fig:AMDCases}
\end{figure}

\subsubsection{Interactions between humans and vectors}

\begin{figure*}[ht!]
\centering
 \includegraphics[trim=1.8cm 15cm 1cm 2.5cm, clip=true, scale=0.8]{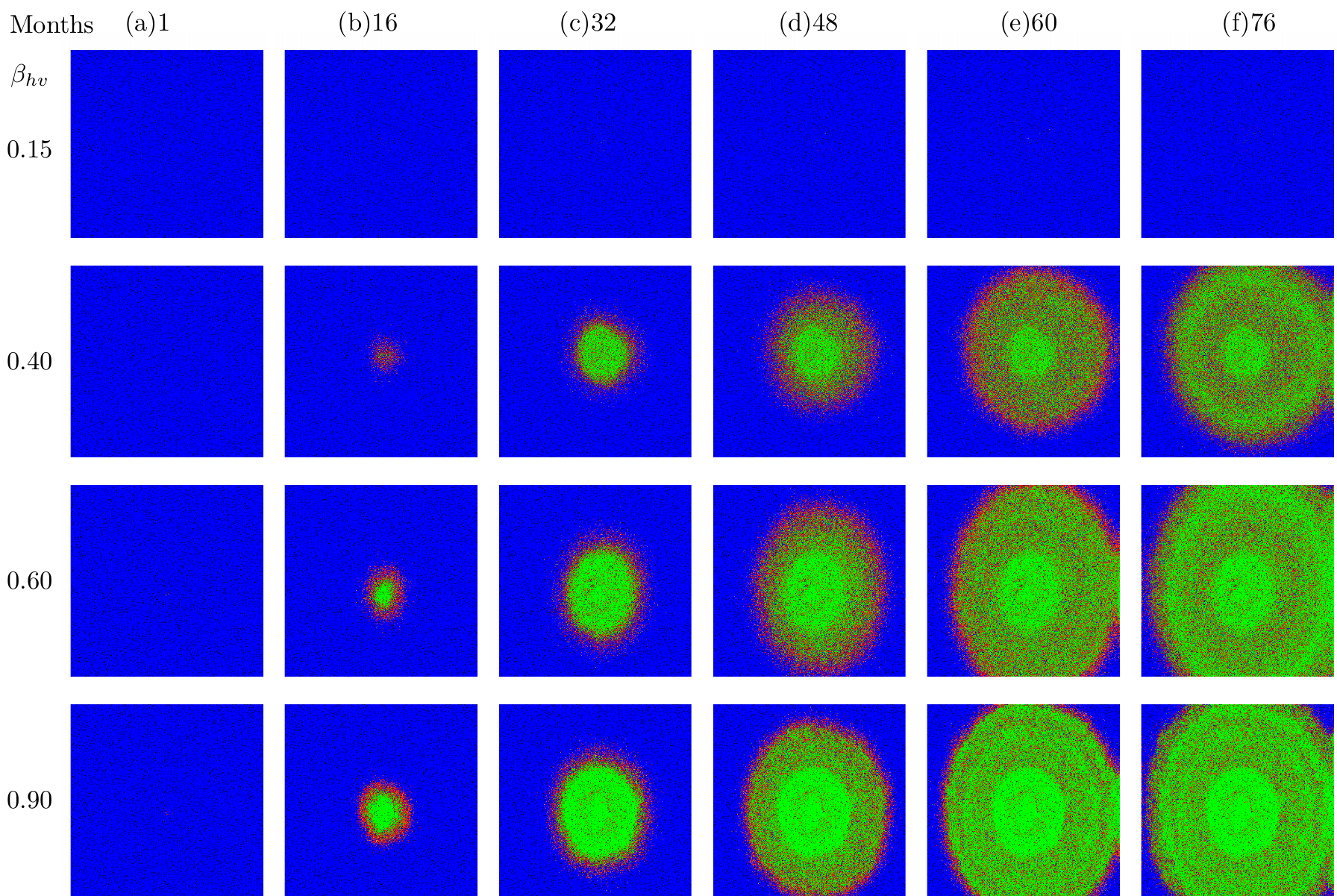}
\caption{(color online) Spatio temporal spread of dengue in stochastic Cellular Automata {\it immobile } humans described in sec. \ref{subsec:VMResults}. 
Columns represents time (months) and rows represents infection transmission rate.  
Coloring scheme depend on the density of S(blue)-E(orange)-I(red)-R(green). }
\label{fig:CAVM}
\end{figure*}

Human-vector and vector-human interaction happens mainly through mosquito bites. We have set the maximum biting rate to two and assumed that the probability of \{0, 1, 2\} bites are \{1/4, 1/2, 1/4\} respectively.  In each cell, an {\it A. Aegepti} mosquito (of any type S,E or I) randomly selects a human to bite and   after each bite, the vector stays in the same cell or follows the vector  mobility pattern described before. The Dengue virus is transmitted from human to vector (or vice-versa) when an infected vector $I_v$ bites a human in a susceptible   state ($S_h$) or an infected human $I_h$ gets bitten by a susceptible mosquito $S_v$ with rate $\beta_{hv}$ $(\beta_{vh})$.

\section{Results}
We have studied two scenarios - immobile and mobile humans, with slightly different objectives. In the immobile case, our focus is to compare the spatio-temporal patterns of Dengue with and without human mobility and test how closely it can explain observed Dengue dynamics in Ahmedabad city (located in the mid-western part of the country) for which Dengue cases and mosquito density data is available from 2006-2012. As an illustration we show Dengue cases in Ahmedabad in 2010 in Fig. \ref{fig:AMDCases}. 
Though our simulations are spatial, we do not include spatial heterogeneties in human and vector density due to lack of data. Our comparison will be limited to temporal variations of our simulation and data. We then describe the results with human mobility, where we highlight the  quantitative differences in fraction of infected and disease extinction time for with and without human mobility.

\subsection{Immobile humans} 
\label{subsec:VMResults}
For the immobile case, we have chosen CA grid size of $L\times L=500\times 500$ and an average human occupancy of 5 agents per cell. This gives us sufficiently big grid population of more than a million to emulate a typical large city in India. Assuming that each cell represents an area $a^2=10m\times 10m=100m^2$, (where $a$  is the width of each cell), we get the model world size to be $25km^2$. In Fig. \ref{fig:CAVM} we show panels of spatio-temporal spread of Dengue in human layer.  The parameters are chosen from Table \ref{tab:ParamTable} with diffusive vector mobility (Section \ref{subsec:vectors}), and  four different values of $\beta_{hv}$ and $\beta_{vh}$ viz. : 0.15, 0.4, 0.6, 0.9.  We choose the following color code: Black represents no human (vector) occupation in the cell. The cell is colored blue at least one person is susceptible.  It is colored  orange or red depending on whether more than 25\% of the population is exposed or infected. In case of conflict, infected color red gets priority. If all persons have recovered, then the cell is colored green. We start the disease dynamics by setting all vectors in the central cell to be infected.
 Based on Fig. \ref{fig:CAVM}, we make the following observations:
 
\begin{figure*}[t!]
\centering
 \includegraphics[trim=2.0cm 21.5cm 1cm 2.5cm, clip=true, scale=0.9]{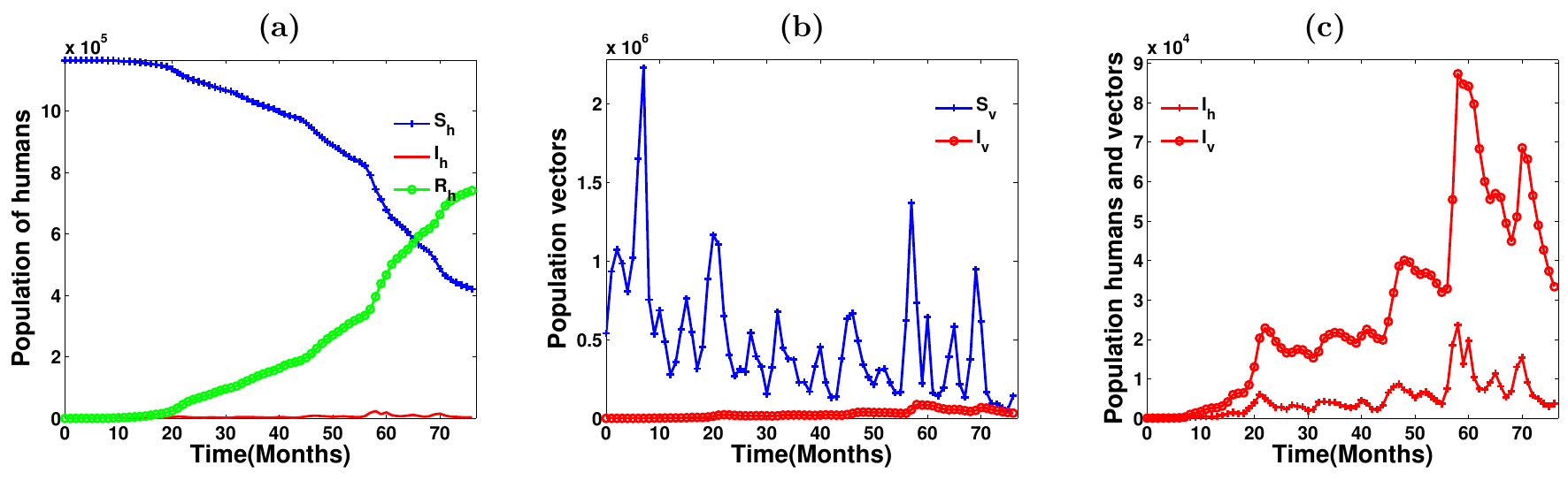}
\caption{(color online) Temporal dynamics of Dengue from stochastic Cellular Automata ($\beta_{hv}=0.4$) 
(a) susceptible, infected and recovered humans  (b) susceptible  and  infected vectors (c) infected humans and infected vectors.   }
  \label{fig:CAVM_temporal}
\end{figure*}

\begin{figure*}[ht!]
\centering
\includegraphics[trim=1.8cm 15.0cm 1cm 2.5cm, clip=true, scale=0.9]{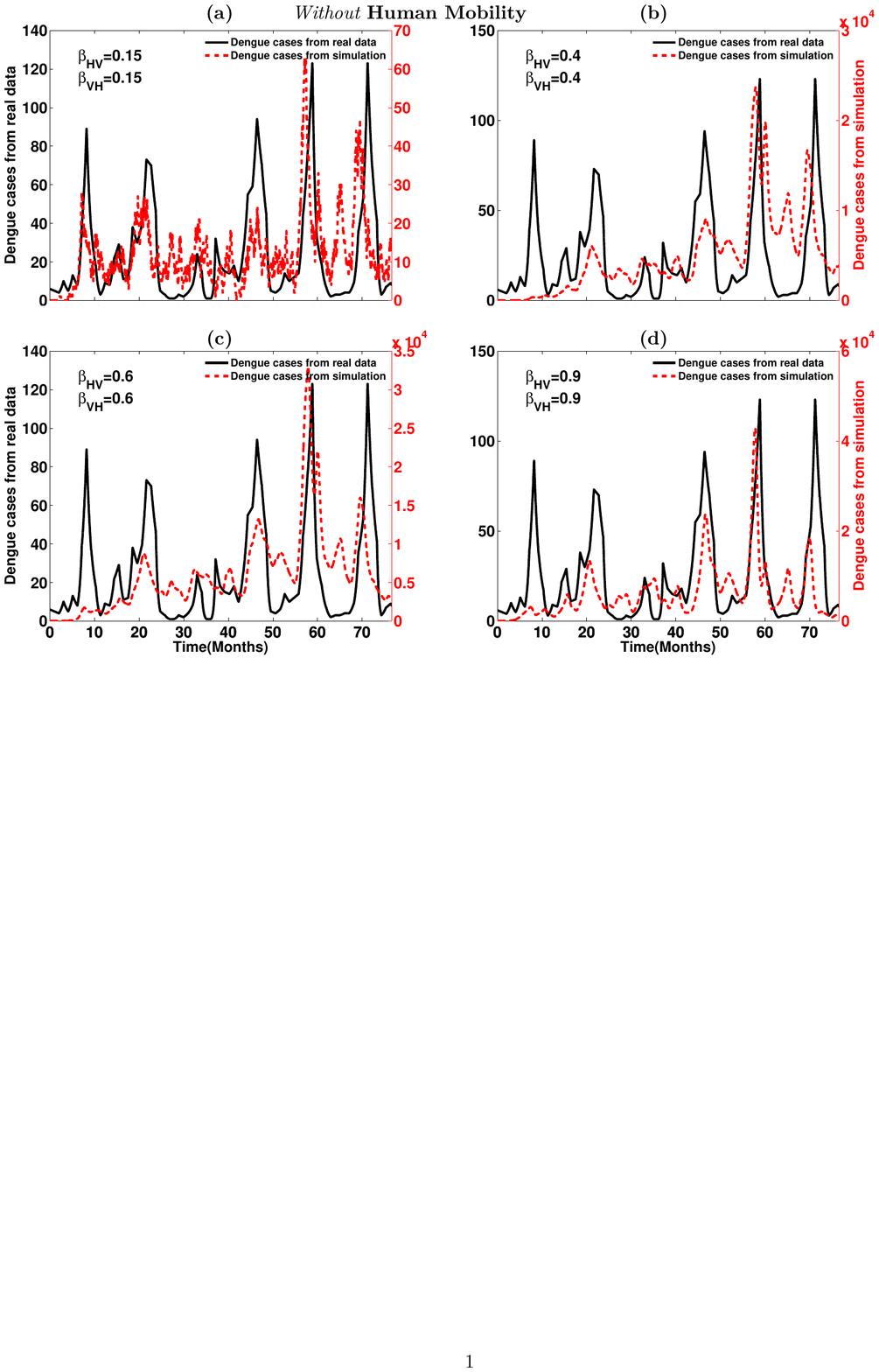}
\caption{(color online) Comparison of Dengue infected people from stochastic Cellular Automata simulation with recorded cases from Ahmedabad city 
from 2006-2012. We have restricted human mobility (a)$\beta_{hv}=0.15$  (b)$\beta_{hv}=0.4$ (c)$\beta_{hv}=0.6$  (d)$\beta_{hv}=0.9$.
(Note the difference in scale in all panels).  }
\label{fig:CAVM_DCCompare1}
\end{figure*}

\begin{figure*}[t!]
\centering
\includegraphics[trim=1.8cm 16.2cm 1cm 2cm, clip=true, scale=0.9]{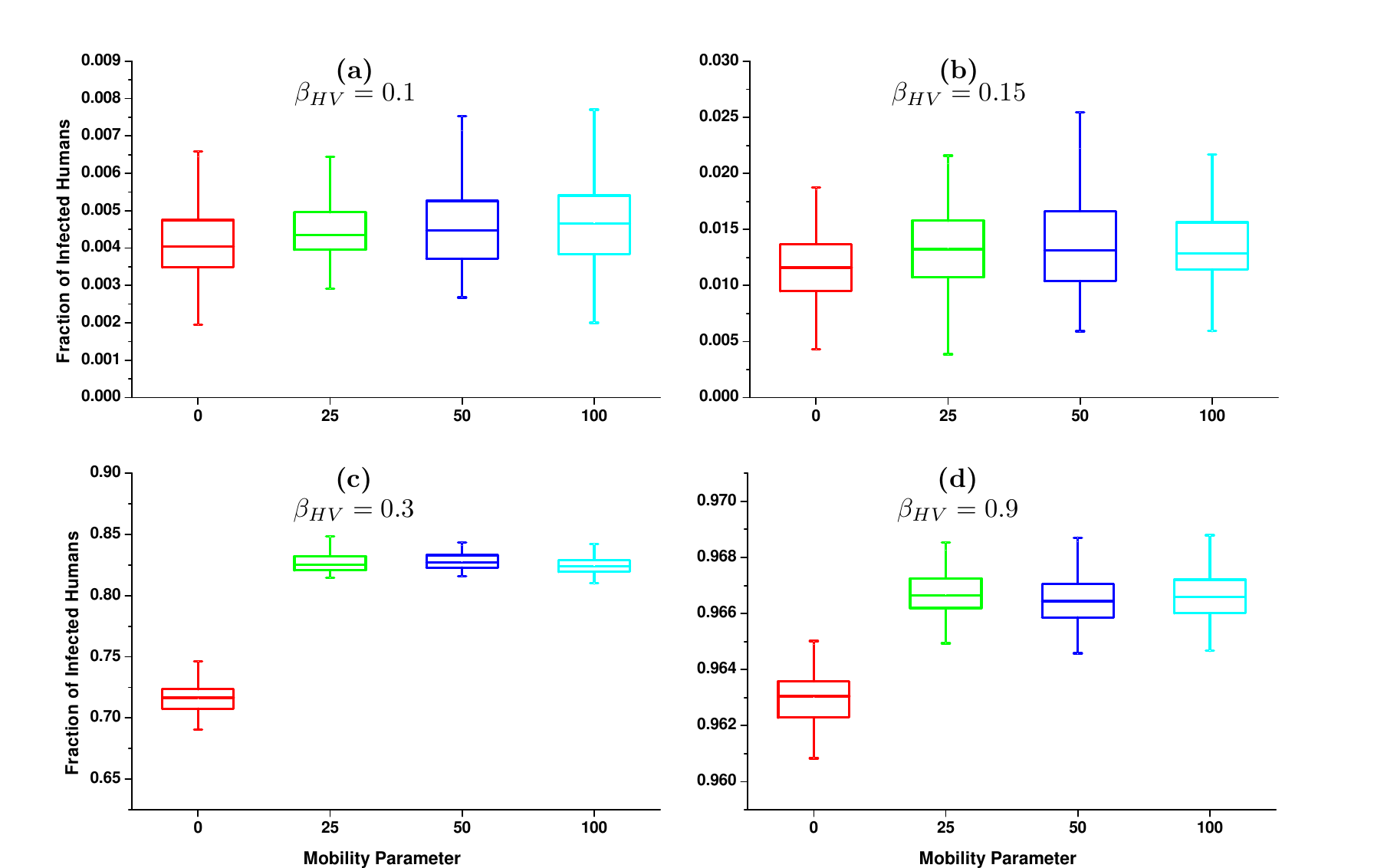}
\caption{Box plots of fraction of infected at the end of simulation with different exponential mobility scale 
($P(l)=1/we^{-l/w}$) $w=\{0,25,50,100\}$ and different infection rate 
(a)$\beta_{hv}=0.1$  (b)$\beta_{hv}=0.15$ (c)$\beta_{hv}=0.3$  (d)$\beta_{hv}=0.9$.}
\label{fig:TotalInf}
\end{figure*}

\begin{itemize}
\setlength{\itemsep}{1pt}
\item At  $\beta_{hv}=0.15$  infection levels are very low and spatial patterns cannot observed. In Fig.\ref{fig:CAVM_DCCompare1}, we see that infections persist at very low levels even after 72 months. But we cannot ascertain distance from the center to which disease goes extinct from the figure. 
 
\item   For higher $\beta_{hv} \ge 0.4$, disease enters the  endemic state and spreads outwards. The precise calculation of transition point is beyond the scope of this work. 

\item Dengue spreads on the human and mosquito lattice in a circular wave  pattern. Vector wave leads the infection . 

\item  In the initial phase, the infected humans are present throughout the inner circle. Later the infected are mainly in the periphery.  In the core, infected humans recover (green) over  time, passing through incubation period. 

\item People in the core who were not infected previously have a risk of infection due to secondary  regeneration of vectors. These periodic outbreaks have high correlation with the temporal variation of the vector density. 

\item In the absence of human mobility, Dengue takes long time to spread to the boundary. Speed of propagation increases with the infection rate.   
\end{itemize}

The temporal patterns of human and vector population, infected human and infected vector and SEI groups of humans  
are shown in Fig. \ref{fig:CAVM_temporal}. Here we observe that at any time, the fraction of infected is quite small comparable to susceptibles (panel a). But, over time large fraction of the population will be infected and  move to recovery phase (as expected from SEIR dynamics, R monotonically increases in active phase if there are no deaths). Similar trend can be seen for vectors in panel b. On a closer look, (panel c), we see that both infected humans and infected vectors show substantial variations in time.
 
In Fig. \ref{fig:CAVM_DCCompare1}, we compare the simulation results for immobile humans (red dash) with the actual Dengue cases in Ahmedabad (black line) for different infection rates $\beta_{hv}(=\beta_{vh})$. For $\beta_{hv}=0.15$, we see that order of magnitude of Dengue cases (max 60 in simulations and 120 in data) match. Peaks increase in strength and become narrower as $\beta_{hv}$ is increased. Peak infection changes by 3 orders of magnitude between $\beta_{hv}=0.15$ and 0.4, where as it increases only by factor 2  between $\beta_{hv}=0.4$ and 0.6, possibly due to core cells being saturated with recovered people who can no longer be infected (see also Fig. \ref{fig:CAVM}). Some peaks seen in the simulations are not present in the data (for example in 30$^{th}$ or 65$^{th}$ month). This may be due to the fact that our data reflect only Dengue cases confirmed  positive in the serum tests conducted by Government laboratory. Data of people with Dengue who have recovered without any serious symptoms is not available.

\begin{figure*}[ht!]
\centering
\includegraphics[trim=1.8cm 16.2cm 1cm 2cm, clip=true, scale=0.9]{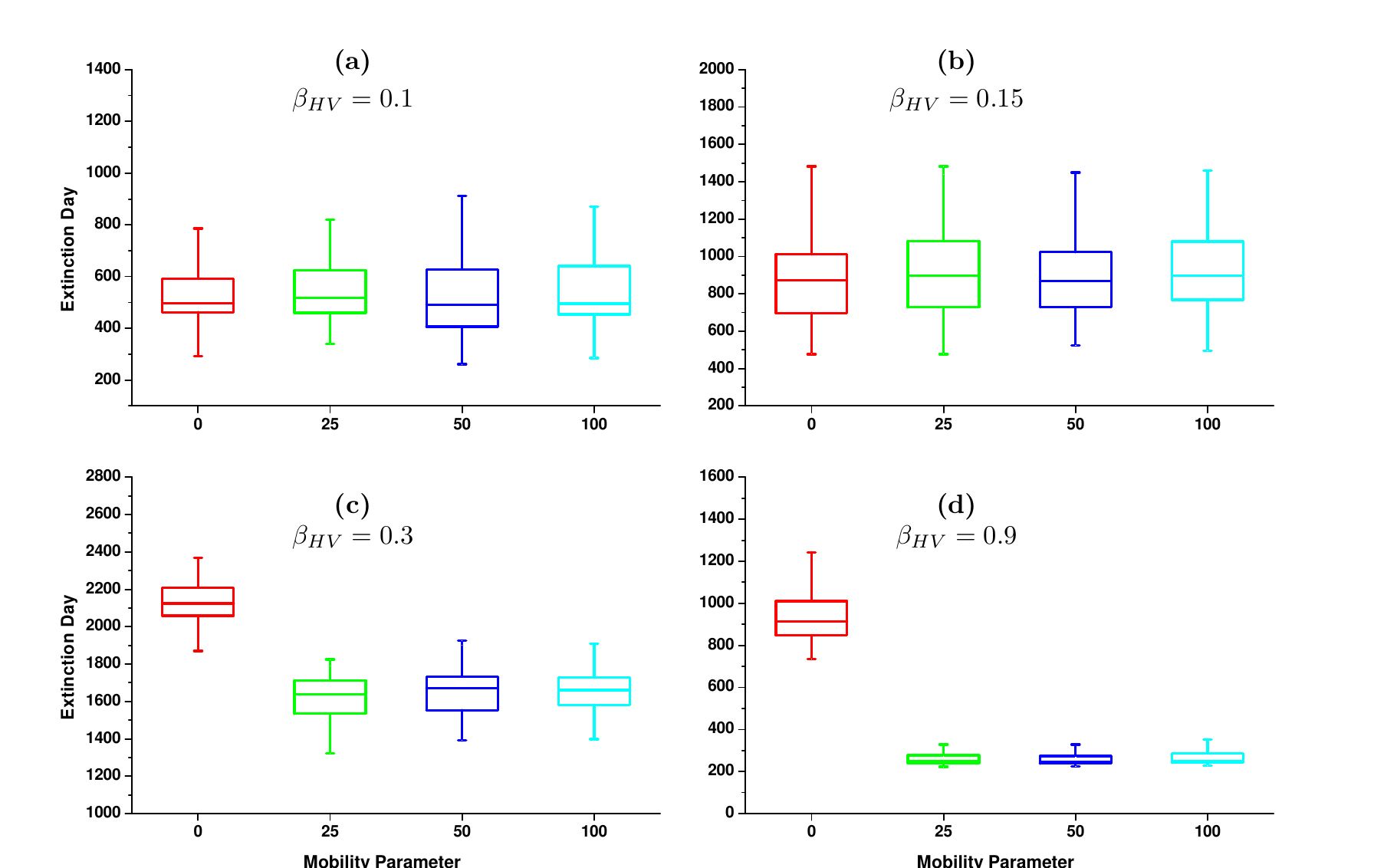}
\caption{Box plots of number of days for extinction with different exponential mobility scale $w=\{0,25,50,100\}$ and different infection rate 
(a)$\beta_{hv}=0.1$  (b)$\beta_{hv}=0.15$ (c)$\beta_{hv}=0.3$  (d)$\beta_{hv}=0.9$.}
\label{fig:ExtinctionDay}
\end{figure*}

\begin{table*}[ht!]
\begin{tabular}{|c|c|c|c|c|}
\hline
$\beta_{hv}$ & w=0        & 25           & 50           & 100          \\ \hline
0.1     & 0.19~(0.18, 0.20) & 0.21~(0.20, 0.22) & 0.21~(0.20, 0.22) & 0.21~(0.20, 0.22) \\ \hline
0.15      & 0.55~(0.52, 0.57) & 0.62~(0.59, 0.65) & 0.63~(0.59, 0.67) & 0.63~(0.60, 0.66) \\ \hhline{|=|=|=|=|=|}
0.3      & 33.35~(33.21, 33.48) & 38.54~(38.46, 38.62) & 38.56~(38.49, 38.63) & 38.41~(38.33, 38.48) \\ \hline
0.4      & 42.15~(42.09, 42.20) & 43.54~(43.49, 43.60) & 43.48~(43.43, 43.54) & 43.38~(43.32, 43.45) \\ \hline
0.6      & 44.27~(44.22, 44.30) & 45.02~(44.97, 45.06) & 45.04~(44.99, 45.08) & 45.07~(45.04, 45.11) \\ \hline
0.9      & 44.89~(44.85, 44.93) & 45.00~(44.96, 45.04) & 45.04~(45.01, 45.08) & 45.02~(44.98, 45.06) \\ \hline
\end{tabular}
\caption{Recovered people (in thousands) at the end of simulation- 95\% confidence Interval about mean 
($\bar{x}\pm 1.96\sigma /\sqrt{n_s}$) (Average population 46.6 thousand and number of simulation runs $n_s$=100)}

\label{tab:Infected}
\end{table*}

\subsection{With human mobility} 

The process of modeling human mobility has been described in Section \ref{subsec:CA}. In this section we describe the results with both vector and human mobility.  Susceptibles can get infected when they move to cells where the vectors in the neighborhood cells are infected. For statistical validation, we choose a smaller lattice $100\times 100$ for all the infection rates and mobility parameters described before, and repeat it 100 times with different random number seeds for statistical averaging. We computed several variables of interest such as (a) total infected population during the entire simulation (recovered at $t=t_f$) (b) disease extinction time (c) time at which infection peaks. 

In Fig. \ref{fig:TotalInf}, we show the box plot of the fraction of total infected population ($R(t_f)/S(0)$) for different mobility 
parameters $w=\{0,25,50,100\}$ at the end of simulation. The $w=0$ (red online) represents immobile case. For low value of $\beta_{hv}$ ($<$0.15), we observe that infection dynamics is in inactive state. For higher $\beta_{hv}$ ($\ge$0.3), disease dynamics enters an active state with large proportion of population  being infected. We can observe (in panel a and b) at low values of $\beta_{hv}=\{0.1, 0.15\}$ fraction of infection population  is not more than 0.02\% for any mobility ($w$). There is no statistical difference between mobile and immobile case and infection levels are fully suppressed. For $\beta_{hv}=\{0.3 - 0.9\}$ (in panel c and d) we can clearly distinguish between fraction of infected population when we include human mobility ($w=\{25,50,100\}$). For very high infection rates (say $\beta_{hv}=0.9$), almost all people are infected, but still small difference is maintained for mobile and immobile cases. In table \ref{tab:Infected}, we have tabulated mean and 95\%confidence interval of $R(t_f)$  for various $w$'s and $\beta$'s.  We observe that in active state, confidence intervals with and without human mobility do not overlap.

Disease dynamics stops when there are no infected humans and vectors. In Fig. \ref{fig:ExtinctionDay} we show the box plot of extinction time for different infection rates and mobility parameters. Panel a and b refers to the absorbing state,  where no statistical difference is seen between for with and without mobility. For the disease endemic state ($\beta\ge 0.3$), we see a surprising result that human mobility has caused disease to die down faster than the case without human mobility.  This result is counter intuitive given that mobility causes greater number of infected people than immobile case (as in Fig. \ref{fig:TotalInf} c and d), but we see disease extinction to be faster when we include mobility. Mobility causes secondary and tertiary waves of infection, who recover in short time and act as barricade for the spreading of primary epidemic wave. Mobility speeds up the infection spread as well as recovery  driving the disease to go extinct earlier. Reaction-diffusion studies in section II A support this finding, where inclusion of mobility (Fig. \ref{fig:pde_h_v} second and third row, $D_h=5D_v,$ $200D_v$) caused early suppression of infection, but not before everyone has recovered.

\section{Summary and outlook}
In this paper we have studied the spatio-temporal dynamics of transmission of Dengue in human and vector population through reaction-diffusion and stochastic Cellular Automata (CA) formalism. We have divided human  and vector population into SEIR-SEI compartments and distributed them on a bi-layer CA lattice.  Coupling between the lattices is through vector bites, which transmits the disease across the layers. Mobility of vectors and humans spread the disease both within and across their lattice. As inputs to the model, we have used parameters from field studies, and vector population data from the Ahmedabad city in India. We have imposed a statistical pattern of human mobility (exponential distribution) on the human lattice to understand how mobility affects the spread of vector borne diseases. For the case without human mobility, we find a good agreement between simulation results of infected humans and data on confirmed Dengue cases in Ahmedabad between 2006-2012. For low $\beta_{hv}$ (0.15), we could reproduce many observed peaks and roughly match the scale. For higher $\beta_{hv}$ many peaks can be reproduced, although their magnitude is much higher. We should note that much of the scale difference might arise due to absence of data on Dengue patients recovering without hospitalization. 

Our main work is to understand the effect of human mobility on Dengue spreading has shown some surprising results. Movement of susceptible and exposed humans lead to an apparent suppression of the epidemic.  Time of extinction of disease is lower when mobility is included, but leads to higher number of people infected during the Dengue cycle. This apparent suppression is primarily due to secondary waves creating recovered regions which block the spread of primary epidemic wave.

In this work we have made several assumptions for simplifying simulations and analysis. Some of these assumptions can be relaxed to give more realistic picture of the Dengue spread. Heterogeneous distribution of population by changing human occupation in lattices based on ward level population density data and excluding regions like rivers and green areas. Exclusion of regions acts as a barrier and prevents secondary epidemic waves emerging from that part. Inclusion of environmental variables such as rainfall patterns can serve as proxy water clogging which acts as breeding sites for vectors. In this work, statistical patterns of human mobility has been taken from studies in US, UK and China. It is not clear (due to lack of literature), whether the same patterns hold good in developing countries such as India, where the current study focuses on. Mobility patterns are closely liked to transportation networks and traffic density. Inclusion of these factors, along with population density can give us more specific insights into how mobility affects spread  of vector borne diseases in cities. We hope that the present work will motivate researchers to take up such studies in the future.

\section*{Acknowledgment}
The authors would like to thank Vinod Reddy, Richard Koblenu, and Profs.  Malavika Subramnyam, Bireswar Das, and Ravindra Amritkar for 
insightful discussions and suggestions. We specially thank Dr. V K Kohli, Assistant Entomologist Ahmedabad Municipal Corporation for 
providing us with data on Dengue incidences and mosquitoes in Ahmedabad city.

\bibliography{references}

\end{document}